\documentclass[a4paper,12pt]{article}
\usepackage[pctex32]{graphics}
\usepackage{amssymb,amsmath}
\usepackage{amsmath,amssymb}
\usepackage{latexsym}
\usepackage{epsfig}
\usepackage[english]{babel}

\newcommand{\be}{\begin{equation}}
\newcommand{\ee}{\end{equation}}
\newcommand{\ba}{\begin{eqnarray}}
\newcommand{\ea}{\end{eqnarray}}

\begin{document}

\begin{titlepage}

\vspace{5mm}

\begin{center}

{\Large \bf Stability of  black holes  in f(R) gravity}

\vskip .6cm

\centerline{\large
 Yun Soo Myung$^{a}$}

\vskip .6cm

{Institute of Basic Science and Department of Computer Simulation,
\\Inje University, Gimhae 621-749, Korea \\}

\end{center}

\begin{center}
\underline{Abstract}
\end{center}

We investigate  the stability of  black holes in the viable model of
$f(R)=R+R^2$ gravity which was known to be the best fit for
inflation. These include Schwarzschild and Kerr black holes. Instead
of studying the fourth-order linearized equation around the black
hole background, we use the corresponding tensor-scalar theory of
the Starobinsky model to perform their stability. The Schwarzschild
black hole is stable, while the Kerr black hole is unstable because
of superradiant instability.\vskip .6cm


\vskip 0.8cm

\vspace{15pt} \baselineskip=18pt

\noindent $^a$ysmyung@inje.ac.kr \\

\thispagestyle{empty}
\end{titlepage}

\newpage
\section{Introduction}

 The $f(R)$
gravity~\cite{NO,sf,FT,Nojiri:2010wj} has  much attentions as a
strong candidate for explaining the current and future accelerating
phases~\cite{SN1,SN2}.   On the other hand, $f(R)$ black holes have
included the Schwarzschild-de Sitter black hole~\cite{CENOZ} and
Schwarzschild-anti de Sitter black
hole~\cite{delaCruzDombriz:2009et}. The trace of energy-momentum
tensor must  be zero to obtain a constant curvature black hole when
$f(R)$ gravity couples with other matters of the Maxwell
field~\cite{delaCruzDombriz:2009et}, the Yang-Mills
field~\cite{Moon:2011hq}, and a nonlinear Maxwell
field~\cite{Sheykhi:2012zz}.

 Interestingly, $f(R)=R+R^2/(6M^2)$
gravity~\cite{Starobinsky:1980te,Whitt:1984pd,Ferrara:2014ima} has
shown  a strong evidence for inflation to support  recent Planck
data~\cite{Ade:2015lrj}.  An important feature of this model
indicates  that the inflationary dynamics were driven by the purely
gravitational sector $R^2$ and the scale of inflation is linked to
the mass parameter $M^2$. However, it cannot work as a successful
model for explaining late-time acceleration.

Most of  astrophysical black holes  are considered  to be a rotating
black hole~\cite{Herdeiro:2014goa}.  The stability analysis of  the
rotating  Kerr black hole is not a routine work as one has performed
the stability analysis of a spherically symmetric Schwarzschild
black hole~\cite{Regge:1957td,Zerilli:1970se,Vishveshwara:1970cc}
because it is an axis-symmetric spinning  black hole.  The Kerr
black hole has been proven to be stable against a massless
graviton~\cite{PT,TP,Whit} and a massless scalar~\cite{DI}. However,
there exist another instability of the superradiance known as the
black-hole bomb when one introduces a massive boson like
scalar~\cite{ZE,CDLY,HH,Dol,Witek:2012tr,Dolan:2012yt} and
vector~\cite{Pani:2012vp}.

It was first noted that the Kerr black hole obtained from  $f(R)=R+h
R^2$ is unstable since it could be transformed into the Brans-Dicke
theory~\cite{Hersh:1985hz}. The Kerr solution could be obtained from
a limited form of $f(R)=a_1R+a_2R^2+a_3R^3+\cdots$
gravity~\cite{kerr}. Importantly, a perturbed Kerr black hole could
distinguish Einstein gravity with two degrees of freedom (DOF) from
$f(R)$ gravity with three DOF~\cite{BS}.  However, it is worth
noting that  the stability analysis of $f(R)$-rotating black hole
was not completely performed because the linearized equation for
$f(R)$ gravity contains fourth-order derivative terms.  One way to
avoid this difficulty is to transform  the limited form of $f(R)$
gravity into a scalar-tensor theory with two auxiliary fields,
leading to that the $f(R)$-rotating black hole is unstable against a
massive scalar perturbation in the Jordan frame~\cite{Myung:2011we}.
Further, the linearized Ricci scalar equation obtained from  the
limited $f(R)$ gravity has shown a superradiant instability if the
linearized Ricci scalar is considered as a massive spin-0 graviton
propagating on the Kerr spacetime~\cite{Myung:2013oca}.

In this work, we wish to focus on performing  the stability of
Schwarzschild and Kerr black holes in the specific model of
$f(R)=R+R^2/(6M^2)$ gravity because its scalar-tensor theory was
clearly shown  to be the Starobinsky model in the Einstein frame
which was extensively investigated as a promising single field
inflation model.   Even though the Starobinsky potential takes the
same form, its role in the black hole physics differs from the
inflation.

\section{$f(R)$ black holes}
We start with  a specific $f(R)$ gravity
\begin{eqnarray}
S_{\rm f}=\frac{1}{2\kappa^2}\int d^4 x\sqrt{-g} f(R),~~
f(R)=R+\frac{R^2}{6M^2}\label{Action}
\end{eqnarray}
with $\kappa^2=8\pi G=1/M^2_{\rm P}$. Here the mass parameter $M^2$
is chosen to be a positive value, which is consistent with the
stability condition of $f''(R)>0$~\cite{sf}. The Einstein equation
takes the form
\begin{eqnarray} \label{equa1}
R_{\mu\nu} f'(R)-\frac{1}{2}g_{\mu\nu}f(R)+
\Big(g_{\mu\nu}\nabla^2-\nabla_{\mu}\nabla_{\nu}\Big)f'(R)=0,
\end{eqnarray}
where the prime (${}^{\prime}$) denotes the differentiation with
respect to its argument.  It is well-known that Eq.(\ref{equa1})
provides the Kerr black hole solution with
$\bar{R}=\bar{R}_{\mu\nu}=0$. Hereafter we denote the background
quantities with the overbar. In this work, we use the
Boyer-Lindquist coordinates to represent an axis-symmetric Kerr
black hole  with mass $\tilde{M}$ and angular momentum
$J$~\cite{Kerrsol}
\begin{eqnarray}
ds^2_{\rm Kerr}&=&\bar{g}_{\mu\nu}dx^\mu dx^\nu= -\Big(
1-\frac{2\tilde{M}r}{\rho^2}\Big)dt^2
-\frac{2\tilde{M}r a\sin^2\theta}{\rho^2}\, 2 dt d\phi  \nonumber \\
& & +\frac{\rho^2}{\Delta}\,dr^2 +\rho^2 d\theta^2+\Big(
r^2+a^2+\frac{2\tilde{M}r
a^2\sin^2\theta}{\rho^2}\Big)\sin^2\theta\, d\phi^2 \, \nonumber \\
& &
 \label{Kerr}
\end{eqnarray}
with
\begin{eqnarray}
\Delta=r^2+a^2-2\tilde{M}r,~ \rho^2=r^2+a^2 \cos^2\theta,~
a=\frac{J}{\tilde{M}}.
 \label{metric parameters}
\end{eqnarray}
In the non-rotating limit of $a\to 0$, (\ref{Kerr}) recovers a
spherically symmetric Schwarzschild
 solution
\begin{equation} \label{Sch}
ds^2_{\rm
Sch}=-\Big(1-\frac{2\tilde{M}}{r}\Big)dt^2+\frac{dr^2}{1-\frac{2\tilde{M}}{r}}+r^2d
\Omega^2_2,
\end{equation}
while the
 limit of $a\to1$ corresponds  to the extremal Kerr black
hole. From  the condition of $\Delta=0(g^{rr}=0)$, we determine two
horizons  which are located at
\begin{equation}
r_\pm=\tilde{M}\pm \sqrt{\tilde{M}^2-a^2}.
\end{equation}
In the non-rotating limit (\ref{Sch}), the  event horizon is given
by
\begin{equation}
r_{\rm EH}=2\tilde{M}.
\end{equation}
 The angular velocity at the outer (event)
horizon takes the form
\begin{equation}
\Omega=\frac{a}{2 \tilde{M }r_+}=\frac{a}{r_+^2+a^2}. \label{hav}
\end{equation}
\section{Black holes in the Starobinsky model}
Since the $f(R)$ gravity provides three DOF, one could represent it
as a scalar-tensor theory  by introducing one auxiliary field
$\psi$~\cite{Whitt:1984pd}
\begin{equation}
S_{\rm A}=\int d^4x\sqrt{-g^J}\Big(\frac{M^2_{\rm
P}}{2}R+\frac{M_{\rm P}}{M}R\psi-3\psi^2\Big),
\end{equation}
where the superscript $J$ means the Jordan frame. Integrating out
the field $\psi$ leads to the original $f(R)$ gravity
(\ref{Action}). Employing the conformal transformation  and
redefining the scalar field ($\psi\to \phi$) to arrive at the
Einstein frame
\begin{equation}
g^J_{\mu\nu}\to e^{-\sqrt{\frac{2}{3}}\frac{\phi}{M_{\rm P}}}
g^E_{\mu\nu}=\frac{1}{1+\frac{2\psi}{MM_{\rm P}}}g^E_{\mu\nu},
\end{equation}
we obtain the Starobinsky model~\cite{Ferrara:2014ima}
\begin{equation}
S_{\rm S}=\int d^4x\sqrt{-g^E}\Big[\frac{M^2_{\rm
P}}{2}R-\frac{1}{2} \partial_\mu \phi\partial^\mu \phi-V(\phi)\Big]
\end{equation}
with the Starobinsky potential (see Figure 1 for its graphical form)
\begin{equation}
V(\phi)=\frac{3M^2_{\rm
P}M^2}{4}\Big(1-e^{-\sqrt{\frac{2}{3}}\frac{\phi}{M_{\rm
P}}}\Big)^2.
\end{equation}
The Einstein and scalar equations are given by
\begin{eqnarray}
\label{s-ein}&&G_{\mu\nu}=\frac{1}{M^2_{\rm
P}}T^{\phi}_{\mu\nu},~~T^\phi_{\mu\nu}=\partial_\mu\phi\partial_\nu\phi-\frac{1}{2}g_{\mu\nu}\Big((\partial\phi)^2+V\Big)
\\
\label{s-s}&&\nabla^2\phi-V'=0,~~V'=\sqrt{\frac{3}{2}} M_{\rm
P}M^2e^{-\sqrt{\frac{2}{3}}\frac{\phi}{M_{\rm
P}}}\Big(1-e^{-\sqrt{\frac{2}{3}}\frac{\phi}{M_{\rm P}}}\Big).
\end{eqnarray}
In the case of $\phi=0(V=0)$, we obtain the Kerr solution
(\ref{Kerr}) and Schwarzschild  solution (\ref{Sch}) to
(\ref{s-ein}) and (\ref{s-s}). We note here that a plateau of
$V\approx \frac{3}{4}M^2_{\rm P}M^2$ appears for $\phi \gg1$, which
was used to define a slow-roll inflation.
\begin{figure*}[t!]
\centering
\includegraphics[width=.5\linewidth,origin=tl]{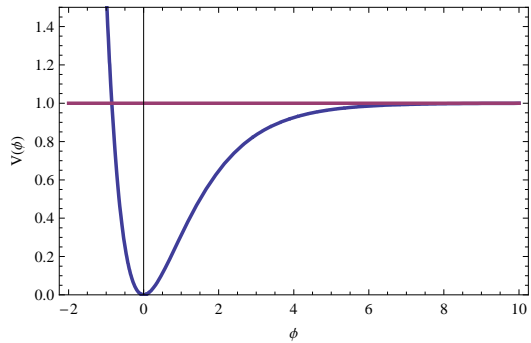}
\caption{Starobinsky potential with $3M^2_{\rm P}M^2/4=1$. In the
case of $\phi=0(V=0)$, we obtain the Kerr and Schwarzschild black
hole solutions, while for $\phi \gg1$ it is sufficiently flat
($V\approx 1$) to ensure slow-roll conditions for inflation in
agreement with the Planck data~\cite{Ade:2015lrj}. }
\end{figure*}
\section{Linearized equations}
 We start with the metric perturbation around the Kerr black hole to
study the linear stability of the black hole
\begin{eqnarray} \label{m-p}
g_{\mu\nu}=\bar{g}_{\mu\nu}+h_{\mu\nu}.
\end{eqnarray}
 The Taylor expansions around $\bar{R}=0$ are  employed  to
define the linearized Ricci scalar~\cite{Myung:2011ih} as
\begin{eqnarray}
f(R)&=& f(0)+f'(0)\delta R(h)+\cdots, \\
f'(R)&=& f'(0)+f''(0)\delta R(h)+\cdots
\end{eqnarray}
with  $f(0)=0$,  $f'(0)=1$, and $f''(0)=1/3M^2$.  The linearized
equation to (\ref{equa1}) is given by
\begin{eqnarray}
\delta R_{\mu\nu}(h)+\frac{1}{3M^2}
\Big[\bar{g}_{\mu\nu}\Big(-\frac{3M^2}{2}+\bar{\nabla}^2\Big)-\bar{\nabla}_{\mu}\bar{\nabla}_{\nu}
\Big]\delta R(h)=0,\label{leq}
\end{eqnarray}
where the linearized Ricci tensor and  scalar are given by
\begin{eqnarray}
&&\delta
R_{\mu\nu}(h)=\frac{1}{2}\Big[\bar{\nabla}^{\rho}\bar{\nabla}_{\mu}h_{\nu\rho}+
\bar{\nabla}^{\rho}\bar{\nabla}_{\nu}h_{\mu\rho}-\bar{\nabla}^2h_{\mu\nu}-\bar{\nabla}_{\mu}
\bar{\nabla}_{\nu}h\Big],\label{lRmunu}\\
&&\delta R(h)=\bar{\nabla}^{\rho}\bar{\nabla}^{\sigma}h_{\rho\sigma}
-\bar{\nabla}^{2}h.\label{lR}
\end{eqnarray}
When using (\ref{lRmunu}) and (\ref{lR}), the linearized equation
(\ref{leq}) becomes  a fourth-order differential equation  with
respect to the metric perturbation $h_{\mu\nu}$. Obviously, it is
not a solvable equation. Another expression for (\ref{leq}) takes
the form
\begin{equation}
\delta
G_{\mu\nu}=\frac{1}{3M^2}\Big(\bar{\nabla}_\mu\bar{\nabla}_\nu-\bar{g}_{\mu\nu}\bar{\nabla}^2\Big)\delta
R
\end{equation}
whose trace equation leads to the linearized Ricci scalar
equation~\cite{Myung:2013oca}
\begin{equation}
 (\bar{\nabla}^2-M^2)\delta R=0.
 \end{equation}

 Choosing the Lorentz gauge of $\bar{\nabla}_\nu
h^{\mu\nu}=\bar{\nabla}^\mu h/2$ and using the trace-reversed
perturbation of $\tilde{h}_{\mu\nu}=h_{\mu\nu}-h
\bar{g}_{\mu\nu}/2$, Eq.(\ref{leq}) takes a  simple from~\cite{BS}
\begin{eqnarray} \label{speq}
\bar{\nabla}^2\tilde{h}_{\mu\nu}+2\bar{R}_{\mu\rho\nu\sigma}\tilde{h}^{\rho\sigma}
-\frac{1}{3M^2}\Big(\bar{g}_{\mu\nu}\bar{\nabla}^2-\bar{\nabla}_{\mu}
\bar{\nabla}_{\nu}\Big)\bar{\nabla}^2\tilde{h}=0.
\end{eqnarray}
Also, one  could not solve (\ref{speq}) directly for
$M^2\not=\infty$ because it is still a fourth-order coupled equation
for $\tilde{h}_{\mu\nu}$ and $\tilde{h}$. However, its trace
equation can be simplified into a factorized form for $\tilde{h}$
\begin{equation}
\label{th-eq}\bar{\nabla}^2\Big(\bar{\nabla}^2-M^2\Big)\tilde{h}=0
\end{equation}
which implies two second-order equations
\begin{eqnarray}
\label{th-eq1}&&\bar{\nabla}^2\tilde{h}=0,\\
\label{th-eq2}&&(\bar{\nabla}^2-M^2)\tilde{h}=0.
\end{eqnarray}

On the other hand, two linearized equations from (\ref{s-ein}) and
(\ref{s-s}) take the simple forms with $\delta T^{\phi}_{\mu\nu}=0$
and $\delta R=0$
\begin{eqnarray}
\label{phi-eq1}&&\delta R_{\mu\nu}(h)=0, \\
\label{phi-eq2}&& (\bar{\nabla}^2-M^2)\varphi=0.
\end{eqnarray}
We note that  two Eqs.(\ref{th-eq2}) and (\ref{phi-eq2}) are the
same but tensor equation (\ref{speq}) [(\ref{leq})] is quite
different from the linearized Einstein equation (\ref{phi-eq1}).
This implies that the complexity of a fourth-order coupled equation
(\ref{leq}) can be reduced to two decoupled second-order equations
(\ref{phi-eq1}) and (\ref{phi-eq2}) if one employs the conformal
transformation and redefinition of scalar field after introducing
the auxiliary formalism, arriving  at  a canonical scalar $\phi$
with the Starobinsky  potential in the Einstein frame. This
describes a process of [$R^2\to R\psi-3\psi^2\to -(\partial
\phi)^2-V$].

\section{Stability of Schwarzschild black hole}

It is a formidable task to perform stability of the non-rotating
Schwarzschild black hole (\ref{Sch}) when one uses the fourth-order
coupled equation (\ref{speq}). Actually, the fourth-order
derivatives appear because the tensor perturbation
$\tilde{h}_{\mu\nu}$ is coupled to its trace $\tilde{h}$. In
addition, the Lorentz gauge condition makes the stability analysis
difficult because one needs to choose the Regge-Wheeler (RW) gauge.
This amounts to double gauge-fixings.  Hence, we must use the
linearized equation (\ref{leq}) to analyze the black hole stability
with the RW gauge. As was mentioned previously, this task is not
available to be performed because Eq.(\ref{leq}) is a fourth-order
differential equation  with respect to  $h_{\mu\nu}$. One way to
perform the stability of the Schwarzschild black hole is to use two
Starobinsky's linearized equations (\ref{phi-eq1}) and
(\ref{phi-eq2}). In this case, it is important to note that the
tensor perturbation $h_{\mu\nu}$ is completely decoupled from the
scalar $\varphi$.   The stability analysis based on (\ref{phi-eq1})
corresponds to that  of the Schwarzschild black hole in Einstein
gravity~\cite{Regge:1957td,Zerilli:1970se,Vishveshwara:1970cc}. It
turned out that the Schwarzschild black hole is stable against the
tensor perturbation. Furthermore, the scalar perturbation based on
(\ref{phi-eq2}) is stable for the mass-squared $M^2 \ge
0$~\cite{Myung:2011ih}.

Consequently,  it  means that the Schwarzschild black hole is stable
against all perturbations in the specific model of
$f(R)=R+R^2/(6M^2)$.

\section{Instability of  Kerr black hole}

 The rotating  (Kerr) black hole has been proven to
be stable against a massless spin-2 graviton~\cite{PT,TP,Whit} and a
massless spin-0 scalar~\cite{DI}. The stability implies that normal
mode solutions were allowed for tensor and scalar propagating on the
Kerr black hole background.  However, there exists another
instability of the superradiance   when one considers  a massive
boson like scalar~\cite{ZE,CDLY,HH,Dol,Witek:2012tr,Dolan:2012yt}
and vector~\cite{Pani:2012vp} around the Kerr black hole background.
We remind the reader that either (\ref{th-eq2}) or (\ref{phi-eq2})
is a massive scalar equation around the Kerr background. Here we
wish to focus on the latter equation. Reminding the axis-symmetric
background (\ref{Kerr}), it is convenient to separate the scalar
into mode~\cite{Teu}
\begin{equation}
\varphi(t,r,\theta,\phi)=e^{-i\omega t + i m \phi} S_{\ell m
}(\theta){\cal R}_{\ell m}(r)\,, \label{sep}
\end{equation}
where $S_{\ell m}(\theta)$ are spheroidal harmonics with $-m\le \ell
\le m$ and ${\cal R}_{\ell m}(r)$ satisfies a radial Teukolsky
equation. Temporary, we may choose a positive frequency $\omega$ of
the mode. Plugging (\ref{sep}) into (\ref{phi-eq2}), one has  the
angular and radial equations for $S_{\ell m}(\theta)$ and ${\cal
R}_{\ell m}(r)$ as
\begin{eqnarray}
&&\hspace*{-2.3em} \frac{1}{\sin \theta}\partial_{\theta}\left (
\sin \theta
\partial_{\theta} S_{\ell m} \right )+ \left [  a^2 (\omega^2-M^2) \cos^2
\theta-\frac{m^2}{\sin ^2{\theta}}+A_{lm} \right ]S_{\ell m} =0,
\label{wave eq separated1}
\\
&&\hspace*{-2.3em} \Delta \partial_r \left ( \Delta \partial_r {\cal
R}_{\ell m}(r) \right )= [\Delta U-K^2] {\cal R}_{\ell m}(r)
 \label{wave eq separated}
\end{eqnarray}
with $U=M^2(r^2+a^2)-2am\omega +A_{lm}$ and $K=\omega(r^2+a^2)-am$.
Here $A_{lm}$ is the separation constant whose form is given
by~\cite{BPT,HH}
\begin{eqnarray}
A_{lm}=l(l+1)+\sum^\infty_{k=1}c_ka^{2k}(M^2-\omega^2)^k
 \label{eigenvalues}
\end{eqnarray}
for $\omega \simeq M$ only. The Teukolsky equation takes the
Schr\"odinger form~\cite{Brill:1972xj}
\begin{equation}
-\frac{d^2\psi}{dr_*^2}+V(r,\omega)\psi=E\psi,~~\psi(r)=\sqrt{\Delta}{\cal
R}(r)
\end{equation}
when the tortoise  coordinate $r_*$ is implemented by $dr_*=
\frac{r^2+a^2}{\Delta}dr$ and $E=\omega^2$. Here, the
$\omega$-dependent potential $V_\omega(r)$ is given by
\begin{eqnarray}
V_\omega(r)=\omega^2+\frac{\Delta U-K^2-\tilde{M}^2+a^2}{\Delta^2}.
\end{eqnarray}
The approximate form of $V_\omega-E$ is given  when keeping
$1/r$-order
\begin{equation}
-E+V^{\rm app}_\omega \to-\omega^2
+M^2-\frac{2\tilde{M}(2\omega^2-M^2)}{r},~~r_*\to
\infty~(r\to\infty),
\end{equation}
while  its form near the event horizon is
\begin{equation}
-E+V_\omega \to (\omega-m\Omega)^2,~~r_*\to -\infty~(r\to r_+).
\end{equation}

\begin{figure*}[t!]
\centering
\includegraphics[width=.6\linewidth,origin=tl]{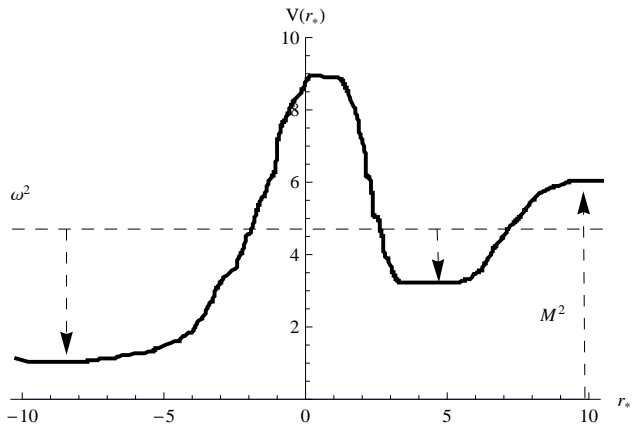}
\caption{Qualitative shape of the Starobinsky potential
$V_\omega(r_*)$. In the limit of $r_* \to \infty$, one finds $V(r_*)
\to M^2$,  a trapping potential well, a potential barrier, and a
potential well in the ergoregion. Here, quasibound states appear for
$\omega^2 <M^2$  because the depth of  potential well in the
ergoregion  is deeper than the depth of trapping potential well.   }
\end{figure*}
At this stage, considering the qualitative  shape of potential
$V_\omega(r_*)$ in Fig. 2 (see Fig.15 of Ref.\cite{KZ} and Fig.7 of
Ref.\cite{Arvanitaki:2010sy}), we could define quasibound states. We
note that the shape of $V_\omega(r_*)$ is slightly different from
$V_\omega(r)$ because $r_*$ goes from $-\infty$ to $\infty$ while
$r\in\{r_+,\infty\}$. For this purpose, we impose the two boundary
conditions of purely ingoing waves near the horizon and a decaying
(bounded) solution at spatial infinity~\cite{ZE}. In this case,
$\omega$ should be complex because flux is absorbed at the horizon.
Near the horizon and at the spatial infinity, the linearized scalar
takes the form~\cite{Dolan:2012yt}
\begin{eqnarray}
\psi_{\{-\infty\}} &\sim & e^{- i(\omega-m\Omega)
r_*}\,\,,\,\,r_* \to -\infty \label{bc1}\\
\psi_{\{\infty\}} &=& e^{-\sqrt{M^2-\omega^2}r_*},~~r_*\rightarrow
\infty. \label{bc2}
\end{eqnarray}
Since the boundary condition at the event horizon is  a purely
ingoing wave, the Kerr black holes do not admit bound states with
real frequency $\omega$.  But, they do admit quasibound states which
have complex $\omega=\omega_R+i\omega_I$ with a negative imaginary
part $(\omega_I<0$), implying  that the decaying field of
$e^{-i\omega_Rt}e^{\omega_It}$  is infalling into the black holes.
For the Kerr black hole, however, there exists  a critical frequency
from (\ref{bc1}): $\omega_R=\omega_c(=m\Omega)$ with $\omega_I=0$,
showing  that there is no scalar  flux into the black hole. For
$\omega_R<\omega_c$, $\omega_I$ becomes positive, implying that the
growing field is falling into black hole. This is the superradiant
regime. It  is a feature of the rotating black hole,  but all
quasibound states on the non-rotating (Schwarzschild) black hole are
found to be decaying. Importantly,  the existence of superradiant
modes can be converted into an instability of the black hole if a
mechanism to trap these modes in a vicinity of the black hole  is
provided.  There are two mechanisms to achieve it. If one surrounds
the black hole by putting a reflecting mirror, the wave will bounce
back and forth between black hole and mirror, amplifying itself each
time and eventually producing a non-negligible backreaction on the
black hole background. It  is not  considered as a perturbation, but
it shows a signal for instability of the black hole.  Secondly, the
nature may provide its own mirror when one introduces a massive
scalar. For $\omega<M$, the mass term works as a mirror effectively.

We remember that any instability must set in via a real frequency
mode and thus, we consider modes with $|\omega_I|\ll \omega_R$ which
implies  $\omega^2\approx \omega^2_R$. From (\ref{bc2}), a bound
state of exponentially decaying mode at spatial infinity is
characterized by the condition
\begin{equation} \label{bc3}
\omega^2<M^2.
\end{equation}
The three boundary conditions (\ref{bc1})-(\ref{bc3}) imply  a
discrete set of resonances $\{\omega_n\}$ which corresponds to bound
states of the linearized  scalar.

More precisely, according to the Hod's argument~\cite{Hod:2012zza},
two conditions are necessary  to trigger the instability of the Kerr
black hole when one considers a massive scalar perturbation: i) The
existence of an ergoregion where superradiant amplification of the
waves takes place.  ii) A trapping potential well
 for quasibound states should exist between the potential
barrier from ergoregion and potential barrier from the mass (see
Fig. 2). The first condition  is  implemented  by the superradiance
condition of $\omega<m\Omega$. The second one is supplied   by  the
condition of the quasibound states for modes in the regime. This
condition can be achieved by considering both (\ref{bc3}) which
states  that $\omega^2$ is less than the potential hight
$V^{\infty}_{\omega}=M^2$ at $r=\infty$ and
 that its approximate derivative must be zero
($dV_\omega^{\rm app}/dr \to 0^+$) as $r\to \infty$. Thus, one has
the condition of
\begin{equation} \label{se-i}
\frac{M^2}{2}<\omega^2<V^{\infty}_{\omega}\to
\frac{M^2}{2}<\omega^2< M^2.
\end{equation}
Combining the superradiance condition with (\ref{se-i}), one finds a
restricted range for the mass
\begin{equation}
M <\sqrt{2} \omega <\sqrt{2} m \Omega
\end{equation}
which implies an  inequality between mass $M$ of the  scalar and
angular velocity $\Omega$ of the Kerr black hole
\begin{equation} \label{ineq}
M <\sqrt{2} m \Omega
\end{equation}
 for the
instability condition of the rotating black hole. On the other hand,
the stability condition is given by
\begin{equation} \label{steq}
M \ge \sqrt{2} m \Omega.
\end{equation}

\section{Discussions}
We have started with  a specific model of $f(R)=R+R^2/(6M^2)$ which
is a fourth-order gravity theory. Considering a process of [$R^2\to
R\psi-3\psi^2\to -(\partial \phi)^2-V$] have led to the Starobinsky
model which is a second-order scalar-tensor theory in the Einstein
frame. This is a famous inflation model.  Even though the stability
of Schwarzschild black hole was not carried out completely within
the perturbed $f(R)$ gravity, its stability was confirmed within the
perturbed Starobinsky model.

In the same spirit, the stability analysis of the Kerr black hole
was performed in the Starobinsky model because the stability
analysis is a formidable task in the framework of the perturbed
$f(R)$ gravity.  The superradiant instability of the Kerr spacetime
 is a consequence of
massive modes that are trapped inside the potential well which
exists between the potential barrier from ergoregion and potential
barrier from the mass at infinity. In this case, quasibound states
appear because the depth of the potential well (near horizon)
outside the ergoregion potential barrier is deeper than the depth of
trapping potential well.

Finally, we wish to mention other models of  $f_p(R)=R+\lambda
R^p$~\cite{Motohashi:2014tra}. The corresponding Starobinsky
potential $V_p$ for $1<p<2$ is steeper than $p=2$ for large $\phi$
(see Fig.1), while $V_p$ for $p>2$ decreases for large $\phi$ and it
approaches zero~\cite{Farakos:2015ksa}. Although the corresponding
potentials have different shapes for large $\phi$, they have the
same behavior around $\phi=0$ with  $V_p(0)=0$  which means that the
two black holes come out as the solution.  However, their stability
analysis seems to be unclear because $\delta V_p'\sim
\varphi^{\frac{p}{p-1}}+\varphi^{\frac{1}{p-1}}$ provides
non-integer power  mass terms. The $p=2$ case of our work leads to
$\delta V'\sim \varphi$.  In other word, we could not obtain a
regular Klein-Gordon equation for $f_p(R)$ gravity. This is closely
related to the fact that $R^p$ for $1<p<2$ could not lead to $\phi$.

\section*{Acknowledgement}

This work was supported supported by the National Research
Foundation of Korea (NRF) grant funded by the Korea government
(MEST) (No.2012-R1A1A2A10040499).

\end{document}